\pdfoutput=1
\documentclass{llncs}
\usepackage{llncsdoc}
\usepackage[usenames]{color}
\usepackage[pdftex]{graphicx}
\usepackage{xspace}
\usepackage{array}
\usepackage{epstopdf}
\usepackage{multirow}
\usepackage{subfigure}
\usepackage{nopageno}
\usepackage{epstopdf}

\newif\ifcomments

\newif\ifanonymize

\def\pname{Redirect2Own\xspace}
\def\ptitle{\pname: Protecting the Intellectual Property 
of User-uploaded Content through Off-site Indirect Access}

\usepackage[hyphens]{url}

\def\eg{{\it e.g.,}\xspace}

\def\ie{{\it i.e.,}\xspace}

\def\fb{Facebook\xspace}

\newcommand{\tcircle}[1]{\textcircled{\scriptsize{#1}}}

\ifcomments
	\newcommand{\gp}[1]{{[[\textcolor{blue}{#1}]]}}
	\newcommand{\angelos}[1]{{\textbullet\textcolor{magenta}{#1}\textbullet}}
\else
	\newcommand{\gp}[1]{}
	\newcommand{\angelos}[1]{}
\fi

\begin{document}

\title{\ptitle}
\ifanonymize
\author{}
\institute{}
\else
\author{Georgios Kontaxis\inst{1} \and Angelos D. Keromytis\inst{2} \and
Georgios Portokalidis\inst{3}}
\institute{Columbia University, New York, NY, USA
\and
Georgia Institute of Technology, Atlanta, GA, USA
\and
Stevens Institute of Technology, Hoboken, NJ, USA
}
\fi
\maketitle

\begin{abstract}
Social networking services have attracted millions of users, including 
individuals, professionals, and companies, that upload massive amounts of 
content, such as text, pictures, and video, every day. Content creators retain 
the intellectual property (IP) rights on the content they share with these 
networks, however, very frequently they implicitly grant them, a sometimes, 
overly broad license to use that content, which enables the services to use it 
in possibly undesirable ways. For instance, Facebook claims a transferable, 
sub-licensable, royalty-free, worldwide license on all user-provided 
content. Professional content creators, like photographers, are particularly 
affected.
In this paper we propose a design for decoupling user data from 
social networking services without any loss of functionality for 
the users. Our design suggests that user data are kept off the 
social networking service, in third parties that enable the 
hosting of user-generated content under terms of service and 
overall environment (\eg a different location) that better suit the user's needs 
and wishes. 
At the same time, indirection schemata are seamlessly integrated 
in the social networking service, without any cooperation from 
the server side necessary, so that users can transparently access 
the off-site data just as they would if hosted in-site. 
We have implemented our design as an extension for the Chrome Web browser, 
called \pname, and show that it incurs negligible overhead on accessing 
``redirected'' content. We offer the extension as free software and its code as 
an open-source project. 
\end{abstract}

\section{Introduction}
\label{sec:intro}

Internet users generate massive amounts of data. Every minute of the day,
smartphones and personal computers are used to share 684,478 pieces of content
on Facebook, add 3,125 photos on Flickr, and upload 48 hours of video on
YouTube~\cite{datapermin:domo12}. Social networking services (SNS) have only
further intensified content creation and dissemination, and maybe more
importantly, they have pushed the number of content creators to new levels.
Facebook, probably the largest such network, reports 1 billion active users as 
of October 2012~\cite{fbstats}. Content creators using such services retain the
intellectual property (IP) rights on the content they share, however very
frequently they implicitly grant the service, a sometimes, overly broad
license to use that content. For instance, according to Facebook's terms of
service (ToS) \cite{fbtos}, the social network gains a \textit{``transferable,
sub-licensable, royalty-free, worldwide license''} on all content the user
uploads or shares with the service. Similarly, Google's ToS informs its 425
million users \cite{googlestats} that \textit{``When you upload or otherwise
submit content to our Services, you give Google (and those we work with)
worldwide license to use, host, store, reproduce, modify, create derivative
works, communicate, publish, publicly perform, public display and distribute
such content''} \cite{googletos}. 


Arguably, the license requested by the various service providers may only
serve the purpose of providing legal support to operations related with data
backups and distributed content delivery. Nonetheless, it \emph{does} include 
uses
that can be in conflict with the IP owner's best interest. For example, photos
or text shared by a user could be used in an undesired way such as being part
of an advertisement campaign targeting their friends or contacts in general.
SNS are also increasingly used by professional artists (\eg photographers) and
businesses for marketing
purposes~\cite{viral_marketing_facebook,fbpages_stores:nytimes12}, and to
promote their intellectual property~\cite{fbpromote}. Such professional
content creators are affected even more profoundly by such ToS. To make things
worse, even if a user removes content from the network, something that
invalidates the granted license, the data is not necessarily immediately
deleted from the server~\cite{fbdelete0}. Note that Facebook actually took steps 
to remedy this situation~\cite{fbdelete1}, but not all services have done so.

\gp{Stretching it a bit.}

At the same time, there are alternative user-content hosting and exchange
services that feature more favorable terms of service to users. For example,
Flickr~\cite{flickr}, Fotki~\cite{fotki}, and Shareapic~\cite{shareapic} are
only a few of the online services that enable the uploading and sharing of
photos under a well-defined license~\cite{flickrtos}, which is limited to the
specific functions necessary to run the offered services (\eg data backup
operations). Dropbox offers similarly favorable ToS~\cite{dropboxtos} to users
for hosting any type of file, while untrusting users can also utilize their
own personal storage device, like  FreedomBox~\cite{freedombox}.


In this paper, we propose a design that decouples content from social
networking services, while still enabling users to share their content through
the social network as before. In other words, \emph{our proposal facilitates
the hosting of user content in third-party, intellectual-property-friendly
hosting services, at the same time allowing its seamless integration in
first-party sites, so that its users can view, like, tag, comment, and
generally interact with it as they normally would}. Our approach employs an
\emph{indirection scheme}, based on replacing the actual content being shared
on the SNS with pseudo-content, containing encoded meta-information that
points to the actual off-site data. This level of indirection places the
original content outside the social network, thereby enabling the IP owner to
share content without relinquishing any rights to the network.

To test our proposal, we implemented an extension for the Chrome Web browser
that performs our indirection design for sharing content on Facebook, in a
transparent way to  users. Our implementation is data specific but not service
specific. We intercept photos being uploaded to Facebook, and instead upload 
them
to a more favorable hosting service of the user's choice (\ie Flickr). After
the upload concludes, we encode the Flickr url\footnote{We can request that
the Flickr url is not indexed by the service, so that the photo is only
accessible to users that know the url. A similar approach is followed by 
Facebook.} pointing to the image in a new pseudo-image, which in turn is 
uploaded to Facebook. The extension can easily modified to also support other 
SNS or sites that store the same type of data, \ie images. We offer our 
extension, called \pname, as free software and have made its code open source. 
Our evaluation of \pname shows that it has negligible overhead; little more than 
simply downloading the image directly from the third-party hosting service.

\vspace{0.6em}
The contributions of this paper can be summarized in the following:
\begin{itemize}
    \item We discuss the terms of service of social networks in 
    respect to the protection of intellectual property in 
    user-provided content.
    
    \item We propose the use of indirection schemata to protect the 
    intellectual property of users in social networks through the 
    off-site hosting of the actual content, thereby placing it 
    outside their legal domain.
    
    \item We offer an implementation of our design as a Chrome 
    extension which can be used for Facebook.
\end{itemize}

The rest of this paper is organized as follows. Section~\ref{sec:related} 
discusses related work. We present the design of our approach in 
Sec.~\ref{sec:design}. We provide implementation details in 
Sec.~\ref{sec:implementation}, and evaluate the performance of \pname in 
Sec.~\ref{sec:evaluation}. Section~\ref{sec:discussion} discusses limitations, 
future extensions, and possible legal implications of our proposal. Finally, 
conclusions are in Sec.~\ref{sec:conclusions}.





\section{Related Work}
\label{sec:related}


FaceCloak~\cite{facecloak} shields a user's personal information from the 
social networking service by providing fake information to the service itself 
and storing the actual, sensitive information in an encrypted form on a separate 
server. 
The authors of FlyByNight \cite{flybynight} propose the use of public key 
cryptography among friends in a social networking service setting so as to 
protect their information from a curious social provider and potential data 
leaks. 
NOYB~\cite{noyb} partitions user information in his online profile into atoms 
and then substitutes each atom with an atom from another user pseudorandomly. 
The mapping of which atoms belong to whom is encrypted. Encryption keys are 
distributed among users authorized to de-scramble a user's profile by placing 
the correct atoms in their place. 
The major drawback of such approaches is the key or secret management 
overhead along with the fact that they hinder the ability of a social 
network user to reach a large unknown audience in a broadcast fashion 
to promote himself (e.g., celebrities) or his work (e.g., artists). 

In the same spirit of data privacy, a number of approaches propose the 
separation of environments, such as social networks, where functionality 
and data are fused together. 
Shakimov et al. \cite{vis-a-vis} address user concerns for the privacy
of their data in social networks by proposing the use of personal virtual
machines, e.g., instances Amazon's EC2 platform, to hold user data while
arbitrating access control. Their approach can be overlayed on top of the
existing social graph of a social networking service but having the service
store IP address pointers to the virtual machines holding the data, thereby
allowing users to remain in their invested platform. 
Similarly Frenzy \cite{frenzy} is a dropbox-powered social network. 
FreedomBox \cite{freedombox} is a community project to develop tiny, low-watt 
computers, also known as ``plug servers'', holding all the user's data, such 
as e-mail, photos, etc., as opposed to having them on the cloud, and running 
free software enabling personal privacy.

\section{\pname Design}
\label{sec:design}

Here we outline our proposal for an off-site-hosting indirection 
schema which enables users to protect their intellectual property 
against social networks and other Web services with restrictive 
terms of use by placing their content outside their legal domain. 
We call our design \pname.

The main pivot of our approach is the decoupling of user data from 
the platform upon which a Web application is founded. We suggest 
that while a user interacts with a first-party Web site, 
\eg Facebook, his data are hosted by a third party. 
For instance, Flickr could be used for storing images and Pastebin for 
text. At the same time, indirection schemata are employed within the 
first-party site to offer transparent access to the off-site data for 
this user and others. 
Creating an abstraction layer between some data-oriented functionality 
of the first-party site (\eg tagging or commenting on photos of one's 
friends) and the actual storage of the corresponding data (here: 
uploaded photographs) offers the flexibility of hosting different 
types of data in different third parties (\eg Flickr), 
subject to different terms of service. Thus, \pname empowers users 
with better control of their online data without suffering service 
migration overhead (\eg move everything to a different social 
networking service) or any serious loss of functionality. 

As our case study we consider the Facebook social networking service. 
The service encourages its members to share a plethora of data, 
including photos, to improve their online experience which stems 
from their interaction and the interaction of their friends with 
that data. In detail, their friends or anyone, depending on their 
privacy settings, can see the images, comment on them, or place 
a labeled frame (tag) around certain people in photos. 
Our design enables users to enjoy the social features offered 
by Facebook while hosting their data off-site, in third-party 
services like Flickr and Dropbox, so that Facebook's 
restrictive terms of service do not apply to the actual content. 
This provides users 
with the flexibility of choosing a third-party content hosting 
service that features less aggressive terms of service. 
Note that the principles behind our design can be applied 
to other scenarios and first-party services as well. 

The owner of content (\eg a photo) uploads it to a third-party 
user-content hosting service and receives back a URL as 
the means of accessing that online content in the future. 
Subsequently, he encodes that URL into a reference object of the same 
type, \eg a pseudo-image. In the next step, the owner uploads that indirection 
object to the first-party site, as he would normally do with the original 
content. The indirection object will assume the same space that the original 
content would occupy (\eg a photo in a Facebook album), if the user chose to 
upload it directly to the first-party site. That reference object acts as an 
indirection schema containing the meta-data necessary to retrieve the 
original content.

In our implementation of \pname as a browser extension, 
presented in Section \ref{sec:implementation}, 
we discuss how the above process can be completely automated 
to a degree that the user does not have to change his workflow; 
he can keep uploading content to the first-party site in the same way as before, 
while our extension will intercept the process and carry out all the steps 
in the background, without the user realizing that something has changed. 

When an indirection object is displayed in the browser, we use the meta-data 
stored within to fetch the original content, and use it to replace the 
indirection object. Our aim is that the process is hidden from the user, who 
does not need to be aware where the displayed content is actually hosted. This 
way users can still access and interact with the content as they would normally 
do.

\begin{figure}[t]
  \centering
  \includegraphics[width=0.9\textwidth]{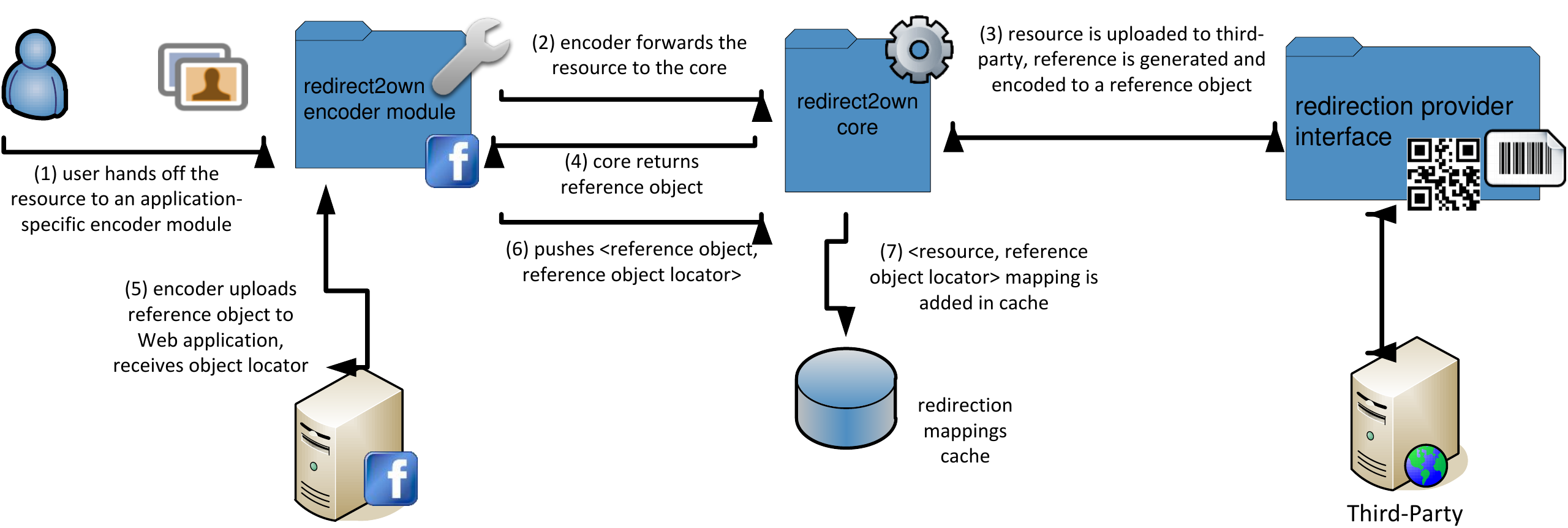}
  \caption{\pname write overview: The process of uploading content 
           to a third-party site, encoding its resource locator into a 
           reference object, and uploading the reference object in the 
           first-party site, taking the place of the content originally 
           uploaded.}
  \label{fig:design_w}
\end{figure}

\begin{figure}[t]
  \centering
  \includegraphics[width=0.9\textwidth]{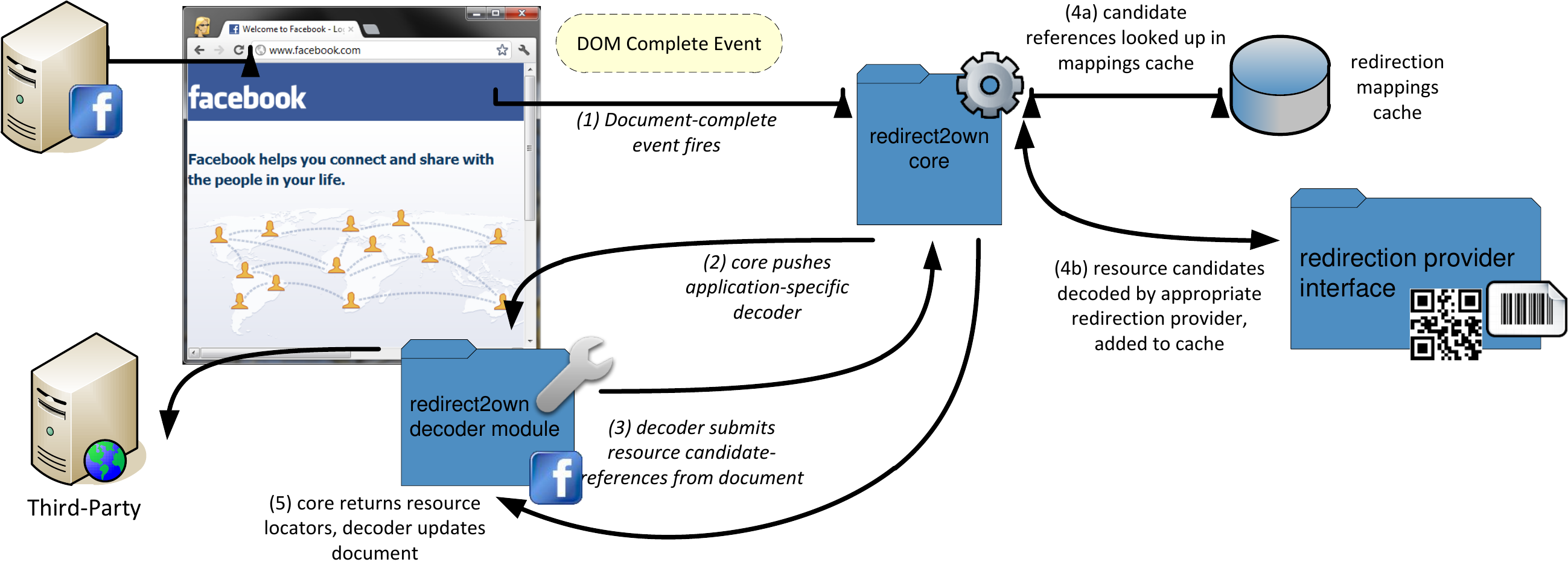}
  \caption{\pname read overview: The process of accessing the 
           original content hosted off-site through its resource 
           locator, obtained by decoding the respective reference 
           object actually hosted by the first-party site.}
  \label{fig:design_r}
\end{figure}

Figures~\ref{fig:design_w} and \ref{fig:design_r} depict the 
proposed design when uploading and accessing content respectively. 
Overall, \pname is organized around three major components; 
the \emph{encoder/decoder}, the \emph{core} and the 
\emph{indirection provider interface}.

The encoder/decoder component carries Web application-specific, 
\eg Facebook, modules that handle \pname's interaction with the 
application when a) the user tries to upload content, and 
b) when the user tries to render a page carrying indirection schemata.
When the user tries to upload (write) content, proper indirection schemata 
(pseudo-data) are presented to the application to take the place of the 
actual data, which are kept off-site in a third-party content hosting 
service. At the same time, the actual content is forwarded to the core 
component which hands it off to the appropriate indirection provider 
module for off-site hosting, in a third party service. 
Then, when the user tries to render (read) content from the first 
party application containing indirection schemata, the decoder module 
is asked by the core component to identify them among different Web page 
elements, so that the core can pass them along to the appropriate 
indirection provider (based on their type) for resolution. Afterwards, 
the decoder module takes care of dynamically replacing the indirections 
with the actual content on the client-side (Web browser) in a manner 
which hides the fact that it is not actually retrieved from within the 
first party application.

The indirection provider carries off-site hosting modules which are 
content-type specific, \eg one or more for storing images and another 
for storing text. These modules handle the hosting of user-generated 
content and return an indirection schema, which is to take the place 
of the original content in the first party service, \eg Facebook. 
These modules also handle the reverse process of resolving an 
indirection to a pointer to the actual content, retrieving the 
content and returning it to the core, which passes it back to 
the decoder module tasked with dynamically embedding it in the 
current Web page at the client-side (Web browser) in the place 
of the indirection object. Examples of indirection provider 
modules are one using Flickr to upload images while encoding their 
URLs inside QR codes in the form of images, and one using Pastebin to 
upload text while encoding the respective URLs using short URLs with 
a custom fragment identifier (\#) at the end to facilitate efficient 
identification by a decoder module.

\subsection{Uploading user-generated content}

We will use the scenario of uploading a photo on \fb and redirecting the content 
to Flickr instead, as an example that will assist us in describing the process 
shown in Fig.~\ref{fig:design_w}.
In step \tcircle{1}, a Facebook-specific encoder module steps in and 
intercepts the uploading of the photo, so that it never 
transmitted from the Web browser toward Facebook. 
Alternatively, the user hands in the photo to the module 
in a manner similar to what he is used to (\ie through a file selection dialog 
windows or by drag-and-drop). 
In step \tcircle{2}, the encoder forwards the photo to the core module,
which picks the appropriate indirection provider based on 
the type of the content (here: image). In step \tcircle{3}, the 
Flickr indirection provider module is invoked to upload 
the photo on the service. Flickr returns a pointer (usually a URL) to the just-
uploaded photo. The indirection module encodes that URL into another image, a QR 
code in this case, as it needs to be able to take the place of the content the 
user intended to upload. Had the original content been something else, \eg text, 
the indirection pseudo-data object would also be in a text form. 
This is to ensure compatibility with the respective container the 
first party service (here: Facebook) utilizes for storing and presenting user 
content. The indirection object (QR image) is returned to the encoder module 
in step \tcircle{4}, which uploads it to Facebook in step \tcircle{5}, and 
acquires a URL reference to it on the Facebook server. This reference is 
used to populate, in steps \tcircle{6} and \tcircle{7}, an indirection mappings 
cache. This cache is used for optimizing indirection object resolution. Parts of 
the cache can also be sent to friends of the user, either out of 
band or through some messaging mechanism Facebook provides. Otherwise, 
users will build their own cache incrementally, as they access the redirection 
objects. 
The process of accessing off-site hosted user content is similar 
and depicted in Figure \ref{fig:design_r}.

%

\subsection{Indirection mappings cache}

The indirection mappings cache maintains the N most frequent 
mappings created, as a result of the user uploading content, 
and the M most recently used mappings, as a result of the 
user rendering indirection schemata that others have created. 
The existence of this cache allows step \tcircle{4b} shown in 
Fig.~\ref{fig:design_w} to be skipped, which can result in reduced, or even 
zero, network requests for fetching an image. 
A hit in the cache saves the downloading and processing of an 
indirection schema (\eg QR-code image) from the first-party 
site. It does not have any impact on the downloading of the 
actual off-site content. However, it can be employed in 
cooperation with the browser's cache, so that frequently 
used content will not be fetched from the network, 
resulting in zero network communication. 
Moreover, mappings from this cache can be exported 
and shared with one's friends in an attempt to 
facilitate the more efficient sharing of content 
collections, such as entire photo albums. Rather 
than having one's friends decode the indirection 
schemata for a newly uploaded collection, communication 
mechanisms offered natively by the first-party service 	
(\eg Facebook's messaging service) can be overloaded 
by \pname to exchange cache mappings between online 
friends. 

\section{\pname Implementation}
\label{sec:implementation}

We have implemented our proposed design in JavaScript (JS)
as a Web browser extension for the latest version of Google Chrome 
(22.0.1229.94 m) 
with encoding modules facilitating uploading images on the Facebook 
social networking service using QR-code images as the indirection 
schemata and \url{imgur.com} as the off-site content hosting service. 
We chose Facebook because of its popularity so that our proof of concept 
implementation supports a representative sample of the user-content 
uploads in social media. We chose \url{imgur.com} as it proved to be 
the fastest service in terms of response time, as shown in Section 
\ref{sec:evaluation}. 
Finally, the decision to adopt QR codes as indirection schemata 
was based on the fact that they are an open standard and any 
QR-capable device can resolve these to the content they point 
to, so that even users who do not have our extension can access 
such content. 

Note that, as described in Section \ref{sec:design}, 
our design is extensible through the addition of modules for 
a plethora of Web applications, including other social 
networking services and third-party content-hosting providers. 


Here we provide implementation details on the components of \pname. 

\subsection{Encoder/decoder component}
\label{ssec:impl_decoder}

As the modules of this component interact directly with Web applications, 
we implemented this as a \texttt{Content Script} \cite{contentscripts} 
running immediately after the \url{window.onload} event 
\cite{window.onload}. 
Content scripts are the front-end of a Chrome extension 
enabling direct interaction with the DOM \cite{dom} of a page. 
The first thing the encoder/decoder does is dynamically load 
application-specific content script modules. These modules register 
filters that specify when each one should run (not all modules run 
for all Web sites) and regular expressions for matching again DOM elements 
that are candidates for carrying indirection schemata (for the decoder). 

To upload content, the encoder component implements a container 
window for the user to drag and drop his image files in. 
Alternatively, the encoder could intercept the user's actions 
in the original upload page within Facebook and carry out the 
steps under the \pname model. We utilize the Facebook Graph 
API (http://developers.facebook.com/docs/reference/api/) to 
upload photos. We then read back the newly-uploaded photo's Graph 
ID and static URL on Facebook, which we add to the mappings cache. 

When browsing Facebook pages, the corresponding content scripts 
are loaded and inspect the DOM once on load and then periodically 
to catch elements dynamically added to the page. 
Once one of the registered expressions matches against a DOM element, 
such as an image, the corresponding content script module is fired.
Matching content is forwarded to the core and then possibly, 
unless the mappings cache lookup returns a hit, 
to the indirection provider component for resolution. 

\subsubsection{Optimizations}
Naively, \pname would attempt to decode every image found in the DOM of a page. 
Even if restricted to pages of a specific domain (\eg Facebook) that would 
waste resources for processing a lot of non-QR images. We optimize the process 
using a few simple heuristics to filter out images that are not part of our 
scheme. First, we filter images based on their location on the server (\eg album 
photos on Facebook are under \texttt{sphotos*.*.fbcdn.net}, while profile photos 
under \texttt{profile.*.fbcdn.net}. By inspecting the dimensions of an image, we 
can also eliminate images that are too small or too large. At the same time, the 
QR-code images that we generate have a fixed 1:1 ratio of their width over 
height, a feature which can help us distinguish them from graphics of similar 
dimensions. We can also ignore image encodings used for other types content in a 
particular service (\eg GIF images).
Finally, if the target service supports adding a text caption to photos  (\fb 
does), we can inject identifiers when uploading pictures to 
facilitate the distinction between QR and non-QR images. 

If multiple QR-codes are encountered on a single page the can be processed in 
parallel using Web Workers \cite{webworkers} in JavaScript. Consequently, those 
images that are indeed valid QR codes containing image URLs, can be also fetched 
in a parallel and asynchronous manner diminishing the network overhead to 
that of a single roundtrip to the off-site content-hosting service.

\subsection{Core component}

The core component coordinates between the front-end components 
(encoder/decoder) and back-end (indirection provider). It is 
implemented as a \texttt{Background Page} \cite{backgroundpages}. 
It dynamically loads the available indirection provider components 
and multiplexes towards them requests, coming from the encoder/decoder 
base, based on the specific Web application and type of content 
being handled. 
It also manages the indirection mappings cache.

\subsection{Indirection Mappings Cache}

As an optimization to resolving indirection schemata to 
actual content each and every time they are requested by 
the user, we employ a mappings cache for speeding up 
frequently used content. This cache takes as input 
the URL of a indirection schema (before it is processed to 
reveal the URL it holds to off-site content -- in our example 
case the URL of the QR-code image hosted on Facebook) and 
returns as output the URL to off-site-hosted content. 
As soon as the decoder component forwards to the core as a 
list of URLs of potential direction schemata, the core does 
a lookup in the mappings cache. On hit, it immediately returns 
the URL to the external content. On miss, it forwards the URLs 
to the indirection provider component. Please note that his 
caching is orthogonal to the Web browser's content cache. 
For frequently-used content, the mappings cache will cause 
the analysis of indirection-candidate objects to be skipped 
and the Web browser's cache will fetch the off-site content 
from the disk rather than the network.

\subsection{Indirection provider component}

We have employed QR codes to implement indirection schemata for images 
and URLs to do the same for text. As our off-site hosting providers, 
we utilize imgur and Pastebin respectively. Both services expose 
a REST HTTP API while its users have to acquire a free registration key. 
This might be an inconvenience for users as it adds one more step 
in the installation process of the \pname browser extension. 
However, other services offer OAuth-based mechanisms which 
smoothen the process. 

At the moment QR codes only contain the URL towards the actual 
resource while we append the \texttt{r2o} suffix in the fragment 
identifier of pastebin URLs so that they can be identified by our 
extension. Both indirection schemata can carry more meta-data 
if necessary. QR codes can encode up to 4,296 alphanumeric 
characters including common symbols and there is no pre-defined 
limit on the length of a URL or its fragment identifier. 


\section{Evaluation}
\label{sec:evaluation}

\pname intercepts content upload and display to transparently implement the 
proposed redirection schemes. Here, we evaluate the overhead in terms of network 
latency and processing time for displaying Web pages containing image-based (\ie 
QR-code images) and text-based redirection schemata. Our evaluation focuses on 
the impact of \pname when \emph{accessing} content through redirection, rather 
than the upload process. We believe that the overhead introduced in content 
display is more crucial for the adoption of our approach, as any piece of data 
is accessed frequently, and generated only once.

\begin{figure}[t]
	\centering
	\includegraphics[width=0.7\columnwidth]{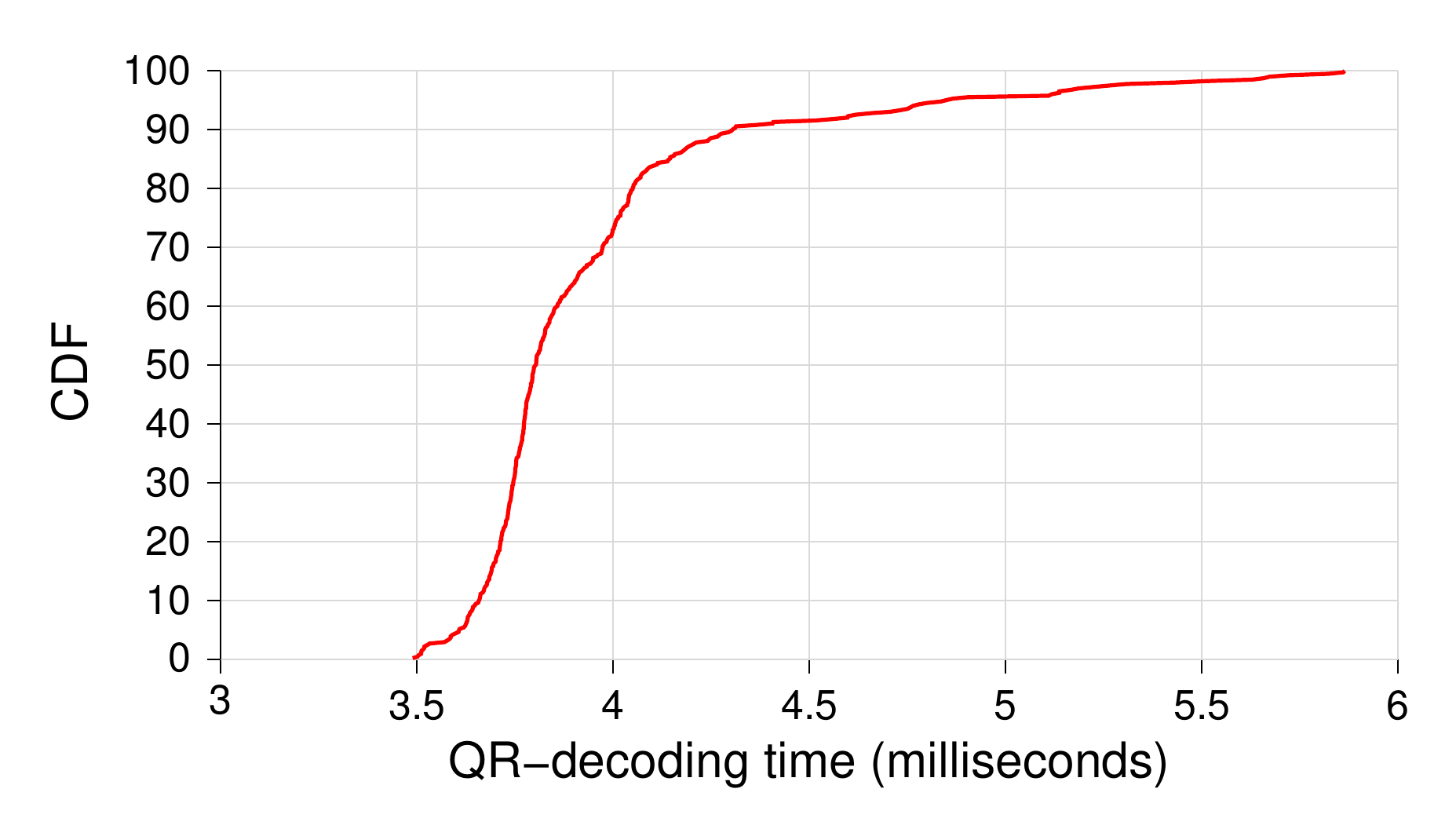}
	\caption{Cumulative distribution function for the time \pname 
	takes to complete a QR-code image-decoding operation. 
	The process will never take more than 6 ms and can be 
	easily parallelized when multiple images are present.}
	\label{fig:cdf_proctime}
\end{figure}

First, we measured the time required for decoding QR-code images. We uploaded 
500 QR codes in a photo album on \fb, and then browsed the albums using Chrome 
(version 22.0.1229.94 m) with \pname enabled. We
instrumented our extension to record the duration of the decoding process. In 
particular, we used the \texttt{getTime()} method of the 
JavaScript \texttt{Date} class to make a note of the time before 
and after decoding a QR code. Because the decoding process exhibited a very 
short duration and our confidence in JS's time reporting was low, we modified 
\pname to decode each QR-code image it encounters 10,000 times, so that the 
process lasts 5 orders of magnitude mode than normal. We then divided the 
obtained results by the number of runs to get the average decoding time for each 
QR code. Figure~\ref{fig:cdf_proctime} presents the cumulative distribution 
function (CDF) for the time it takes to complete a decoding operation. Overall, 
the process of decoding a QR-code image takes on average 4 ms. Bear in mind that 
this overhead is only incurred for images that have not already been filtered 
out by the decoder (see optimizations in Sec.~\ref{ssec:impl_decoder}).

To measure the network delay for retrieving the actual (hosted off-site) content after the resolution of an indirection schema, we employed 8 popular 
image-hosting services. We used the top five results when querying for ``image
hosting'' on Google, plus Flickr, Dropbox, and Facebook's content distribution
network (CDN). For each service, we uploaded a 44KB image once, and then
retrieved it 10 sequential times in 60 second intervals for a period of 32
hours on a weekday. Table~\ref{tab:service_lat} presents the median values
from our measurements. We notice that we can retrieve an image from Facebook's 
CDN (where the Facebook-uploaded images are normally stored) in 11 ms, and only one other service, \texttt{imgur.com}, is equally fast. Flickr takes close to 150 ms, while Dropbox requires almost 300 ms to serve the content. We see a significant gap between the first and last four services. We believe that 
image-hosting services that are more oriented towards interoperability with social networking services, such as Twitter and Facebook, are geared towards offering lower response times than services that advertise ample storage 
for hosting one's images on the Cloud.

\begin{table}[t]
	\centering
	\begin{tabular}{|p{80pt}r|}
		\hline
		{\bf Service}	&	{\bf Response Time}	\\
		\hline
		Facebook CDN & 11 ms \\
		\hline
		Imgur & 12 ms \\
		\hline
		Photobucket & 50 ms \\
		\hline
		PostImage & 51 ms \\
		\hline
		Flickr & 147 ms \\
		\hline
		Dropbox & 306 ms \\
		\hline
		Tinypic & 310 ms \\
		\hline
		Imageshack & 434 ms \\
		\hline
	\end{tabular}
	\caption{Time needed to retrieve a 44KB image from 8 popular image-hosting 
	services.}
	\label{tab:service_lat}
\end{table}

Overall, to access a single image using our indirection scheme, the 
browser would have to download that pseudo-content from the server,
decode it, and then retrieve the actual content from a third-party server, such 
as Flickr. Therefore, in the case of \fb someone would need 162 ms and 321 ms 
for using \pname with Flickr and Dropbox respectively, as opposed to 11 ms had 
the content been posted directly. As we have shown here, that is mainly due to 
the slower response times of these services, and less an effect of the 
redirection scheme.

\subsubsection{Caching Effects}
Besides the fact that multiple schemata can be processed 
and retrieved in \emph{in parallel}, \pname includes a mappings cache for frequently-accessed images or for a collection of images that the owner wants to share with his friends (described in Sec.~\ref{sec:implementation}).
In the first case, we maintain a mapping between the URL of the indirection 
schema on the first-party server (here: Facebook) and the URL of the actual 
content in the third-party server (\eg Flickr). URLs of indirection schemata found in the cache will \emph{not} be retrieved or decoded, and the off-site content will be immediately fetched. Therefore, a cache hit completely eliminates the cost of downloading and decoding a QR code from \fb, \ie 16 ms in total. Please note that this caching method can be used along (\ie it is orthogonal) with the browser's own caching mechanism so that if the off-site content is already cached from a previous request, the overall network delay will remain zero. In the second case, the owner of the content can share the mappings with his friends either out of band or using, for instance, Facebook's messaging system. We imagine this in a scenario, where users would want to share an entire album of photos. This sharing of mappings can be performed transparently to users by \pname, piggybacking on existing social features.






\section{Discussion}
\label{sec:discussion}

\subsection{\pname without a browser extension} 

We consider the case of users accessing off-site content 
in a practical way without the use of the \pname 
browser extension. Enabling such access to the content 
in a practical way allows one's online friends, who cannot 
or do not want to install this extension, to still be able 
to access the content from within the first-party site. 
Such compatibility for non-users of the extension 
is expected to encourage its adoption. 
We thus focus on compensating for the \emph{read} 
functionality of the \pname extension. 
On that matter we observe that social networking 
services, to improve the interaction of their 
users with external sources, offer remote-content 
preview features. In other words, when users post 
links or other references to external content, 
a preview of the remote resource (\eg Web page, 
image or video) is rendered within the social 
networking service in the form of a thumbnail 
picture. The external content, along with the 
preview rendering are not hosted on the social 
networking service, and when users attempt to 
interact with them, they are redirected to 
the original external source. 
They are therefore not part of the social 
networking service and are not covered by 
its terms of service. 

By employing the ubiquitous feature of commenting on 
items featured in a social networking service, we can 
include links or other references, that the service 
will generate a content preview for, to all the 
reference objects. For instance, one can leave a 
comment in Facebook underneath a QR-image reference 
object (generated by \pname) with the URL 
referencing the original image in the third-party 
site. Facebook will render a preview 
thumbnail picture of the original image, based 
on that URL in the comment, enabling users without 
the \pname browser extension to take a glimpse of the content 
being referenced, and  also provide a link for them to easily navigate to the 
hosting service. Such functionality is not odd and 
can be met in other social networking services such 
as Twitter. Besides the use of comments, other 
approaches are through social plugins, \eg 
Facebook's Like and Share buttons that have 
the same result for pages on external sites 
where the original user content might be hosted.

\subsection{Absence of user data from server-side computations} 
Obviously the social networking service is unable to carry out 
any kind of computations on the off-site data. For instance, 
Facebook has rolled out a feature enabling labeled frames (tags) 
of people in photos so that online friends are notified when 
someone they know is ``tagged'' in a new photo uploaded in 
their social graph. If all photos are kept off-site, 
tagging may not work out of the box. 

One may notice that tagging takes place dynamically at the client 
side through code running in the user's browser and the service's 
back end is employed only as a permanent store for that information. 
Therefore, a code bridge could be implemented so that user's tagging 
actions on the off-site data can be translated to valid information 
for the social network to store and serve. We have experimented 
with the simple case of adding a white border around the actual 
QR-code to create a image of the same size as the one hosted 
off-site. We found that tagging works as far as the users are concerned. 

On the other hand, other features such as social authentication 
might be more challenging to support. In social authentication, 
a user is presented with photos of his friends and is prompted 
by the social network to correctly identify them as an additional 
authentication feature besides providing his password. Facebook 
employs facial recognition algorithms to select good-quality 
photos of people's faces for the social authentication challenges. 
Storing indirection schemata instead of the actual photos in 
the social network, will prevent it from generating social 
authentication challenges.

In general, storing user-generated content anywhere but the first-party 
Web site results in any kind of server-side calculation to be infeasible. 
For instance, when storing text, indexing and search functionalities 
cannot be supported by the server as the actual text is retrieved 
by each user on his side and is never available to the server. 
Same thing applies to photos are a more specific case detailed above. 
We consider the calculation of content meta-data at the client side, 
as a means of providing the necessary information for features 
such as the aforementioned to be provided by the first-party site. 

%

\subsection{Legal right to follow links}

There is a special mention in the Facebook terms of service for 
social plugins such as the Like and Share buttons which enable 
users of third-party sites to post links or content from these 
sites to Facebook. The terms of service state that when such 
sites include the plugins in their pages they give permission 
to Facebook to ``use and allow others to use such links and 
content on Facebook''. It is not clear what falls under the 
``use'' of the content coming from external sites. Facebook 
creates an entry in the user's profile after he has ``liked'' 
or shared external content. That entry is called an ``activity'' 
and shared with the user's online friends. 
For that matter, Facebook employs a crawler to fetch one of the 
images from the external page to use it as a thumbnail and also 
inspects the content of the external page to identify keywords, 
categorize it and extract a subset of the text to use as the 
description of this entry. The same behavior applies to URLs 
the user posts in his profile as part of status updates or 
in his comments to the activity or content of other users. 

At the moment there seems to be a technical distinction between 
content, including text and images, retrieved that way from 
external sites and content the user directly uploads to Facebook, 
\eg images added to his photo albums. The former are stored 
under the \url{external} subdomain of Facebook's CDN domain 
(\url{external.ak.fbcdn.net}) while the latter are found 
under the \url{photos} or \url{profile} subdomains 
(\url{http://sphotos-e.ak.fbcdn.net}, 
\url{http://profile.ak.fbcdn.net}). Furthermore anything 
under the external subdomain is periodically fetched from 
its original source, usually after a user's action to 
access it inside Facebook. 

Overall, we cannot find a clear license claim on external content 
and it seems for the moment that there is some distinction between 
that and content uploaded explicitly by users. However, the line 
gets blurry when users post links or content from external sources 
which are part of their own intellectual property. Although these 
sources will be treated as external in the way they are stored, one 
could argue that legally they fall under clause 2 of the terms of 
service \cite{fbtos}, describing the implicit sub-licensing of such 
content posted ``on or in connection with Facebook''.

\section{Conclusions}
\label{sec:conclusions}

Massive amounts of content are generated and shared through social networking 
services every day. Users, be it for leisure or business, upload their content 
online, many times agreeing to terms of service that grant broad licenses on how 
services can use their content and may not meet the users' needs and wishes. We proposed a design that decouples content from 
such services, enabling users to share their content in the first-party service as they normally would, while actually hosting it in a more suitable third-party hosting provider. For example, services with more favorable terms like Dropbox can facilitate such user needs.  
We created \pname, an extension for the Chrome Web browser, that enables users 
to share photos on \fb, while hosting the actual content on Flickr, in a 
transparent and effective way. This extension provides a way for individuals and 
businesses to keep using social networking sites, without relinquishing any use 
rights of their intellectual property. 
Finally, we believe that \pname can also be perceived as the 
means to exercise pressure on such services to amend their terms of service in 
the users' favor.

%

{\small
\bibliographystyle{splncs03}
\pdfbookmark[1]{References}{refs}
\bibliography{photo_ownership}}

\end{document}